\documentclass[twocolumn,fleqn]{article} \usepackage{amsmath,amssymb,amsfonts}
\usepackage{graphicx} 			
\usepackage{url}
\usepackage{siunitx} 
\title {Analysis of COVID-19 infection waves in Tokyo by Avrami equation} 
\author {Yoshihiko Takase \thanks {Chiba JICA Senior Volunteers Association}} 

\begin{document} 
\maketitle 

\abstract {}
The purpose of this study was to simulate all COVID-19 infection waves in Tokyo, the capital of Japan, by phase transformation dynamics theory, and to quantitatively analyze the detailed structure of the waveform for estimating the cause. The whole infection wave in Tokyo was basically expressed by the superposition of the 5 major waves as in Japan as a whole. Among these waves, the detailed structure was seen in the 3rd and the 5th waves, where the number of infections increased remarkably due to the holidays. It was characterized as "New Year holiday effect" for New Year holidays, "Tokyo Olympics holiday effect" for Olympics holidays, and "Delta variant effect" for the replacement of the virus with Delta variant. Since this method had high simulation accuracy for the cumulative number of infections, it was effective in estimating the cause, the number of infections in the near future and the vaccination effect by quantitative analysis of the detailed structure of the waveform.

\section {Introduction}

We have been attempting to analyze the wave of COVID-19 in Japan\cite{sim_covid19_1_japan,sim_covid19_1to5_japan} by the Avrami (or JMAK) equation\cite{avrami} that describes the phase transformation dynamics and is another approach than the conventional SIR model and its expanded models\cite{sir_model1,sir_model2,sir_model3,sir_model4,sir_model5}. As its entity model, the physical process of the linear growth of nuclei forming randomly in the parent phase was assumed. This process is quantitatively determined by three parameters, the initial susceptible $D_\mathrm{s}$, the domain growth rate $K$, and the nucleation decay constant $\nu$. 

The least squares method was applied to the difference between the detected and the theoretical values of the number of time-dependent cumulative daily new infections $D(t)$ (hereinafter referred to as the total infections), and the three parameters, $D_\mathrm{s}$, $K^2$ and $\nu$ for the 2-dimensional linear growth were determined. Since the accuracy of the simulated $D(t)$ was as good as $95\% CI / D(t) < 2.5\%$ after the 2nd wave when $D(t)$ exceeded 10,000 in Japan, it was effective in predicting the number of infections in the near future and in estimating the cause of the detailed structure of the waveform\cite{sim_covid19_1to5_japan}. 

On the other hand, it was difficult to evaluate the error of the obtained three parameters. This is because the Avrami equation is a nonlinear function, and applying the least squares method to $D(t)$ does not always give a unique solution. $D_\mathrm{s}$ was determined accurately because it was a detected value. It was difficult to evaluate the error of $K^2$ and $\nu$ because they were not directly detected values even if they were determined by the least squares method for $D(t)$. Since $K^2$ and $\nu$ had a somewhat complementary relationship, for example, if the value of $K^2$ was increased a little and the value of $\nu$ was decreased a little, the change in the error of $D(t)$ was small in the result of applying the least squares method to $D(t)$. This time, in order to reduce the error factor, the least squares method is applied to $D(t)$ assuming that the value of $\nu$ is constant.

The total infections of COVID-19 in Tokyo prefecture, the capital of Japan, accounts for about 22\% of that in Japan\cite{metro_tokyo,ministry_hlw}. It is important to analyze the infection wave in Tokyo because the spread of infection in Japan often starts in Tokyo, where there is a lot of traffic to and from foreign countries. This time, all COVID-19 infection waves in Tokyo were simulated by the same method as in Japan as a whole\cite{sim_covid19_1to5_japan}. The whole infection wave in Tokyo was basically expressed by the superposition of the 5 major waves as in Japan as a whole. Among these waves, a significant detailed structure was seen specifically in the 3rd and the 5th waves. Both waves showed that the number of infections increased rapidly after the consecutive holidays. 

The 5th wave, which lasted from early June to early October 2021, was the largest infection wave. During this period, the virus was replaced by the more infectious Delta variant\cite{voc_who}. The "Tokyo 2020 Olympic Games", which was postponed for one year, was held in Tokyo. In addition, vaccination progressed during this period.
It will be the most important task to quantitatively analyze the detailed structure of the waveform of the 5th wave and to know the relationship with these events. However, since this study is limited to analysis using physical models, countermeasures in medicine and medical fields and traditional SIR models will not be examined.

The purpose of this study is to simulate all COVID-19 infection waves in Tokyo by the phase transformation dynamics theory, and to quantitatively analyze the detailed structure of the waveform for estimating the cause.

\section {Data analysis} 
\subsection {Theoretical basis}

In the works so far\cite{sim_covid19_1_japan,sim_covid19_1to5_japan}, the least squares method has been applied to $D(t)$ to independently determine the three parameters. This time, for the reason mentioned in the introduction, the least squares method is applied to $D(t)$ assuming that the value of $\nu$ is constant ($\nu = 0.0090$).

The number of time-dependent daily new infections $J(t)$ (hereinafter referred to as the daily infections) is\cite{sim_covid19_1_japan,sim_covid19_1to5_japan} 

\begin {equation} 
\label {eq:avrami_j} 
  J(t) = D_\mathrm{s} \left( -\frac{K^2}{\nu} f_1 \right) \exp\left(-\frac{K^2}{\nu^2} f_2 \right), 
\end {equation} 

where $t$ is the time,  $D_\mathrm{s}$ is the initial susceptible, $K^2 = 2\pi G^2 l_c N_0$, $G$ the growth speed, $l_c$ the domain thickness, $N_0$ the number of active points for nucleus, $\nu$ the decay constant, $f_1 = 1 - \mathrm{e}^{-\nu t} - \nu t$ and $f_2 = 1 - \mathrm{e}^{-\nu t} - \nu t + (\nu t)^2 / 2$.

The number of total infections is expressed by the following equation\cite{sim_covid19_1_japan,sim_covid19_1to5_japan},

\begin {equation} 
\label {eq:avrami_d} 
  D(t) = D_\mathrm{s} \left[ 1- \exp \left ( -\frac{K^2} {\nu^2} f_2 \right ) \right ]. 
\end {equation} 

\subsection {Analysis of the entire waveform}

\begin {figure} [p] 
	\centering 
	\includegraphics[width=8.0cm]{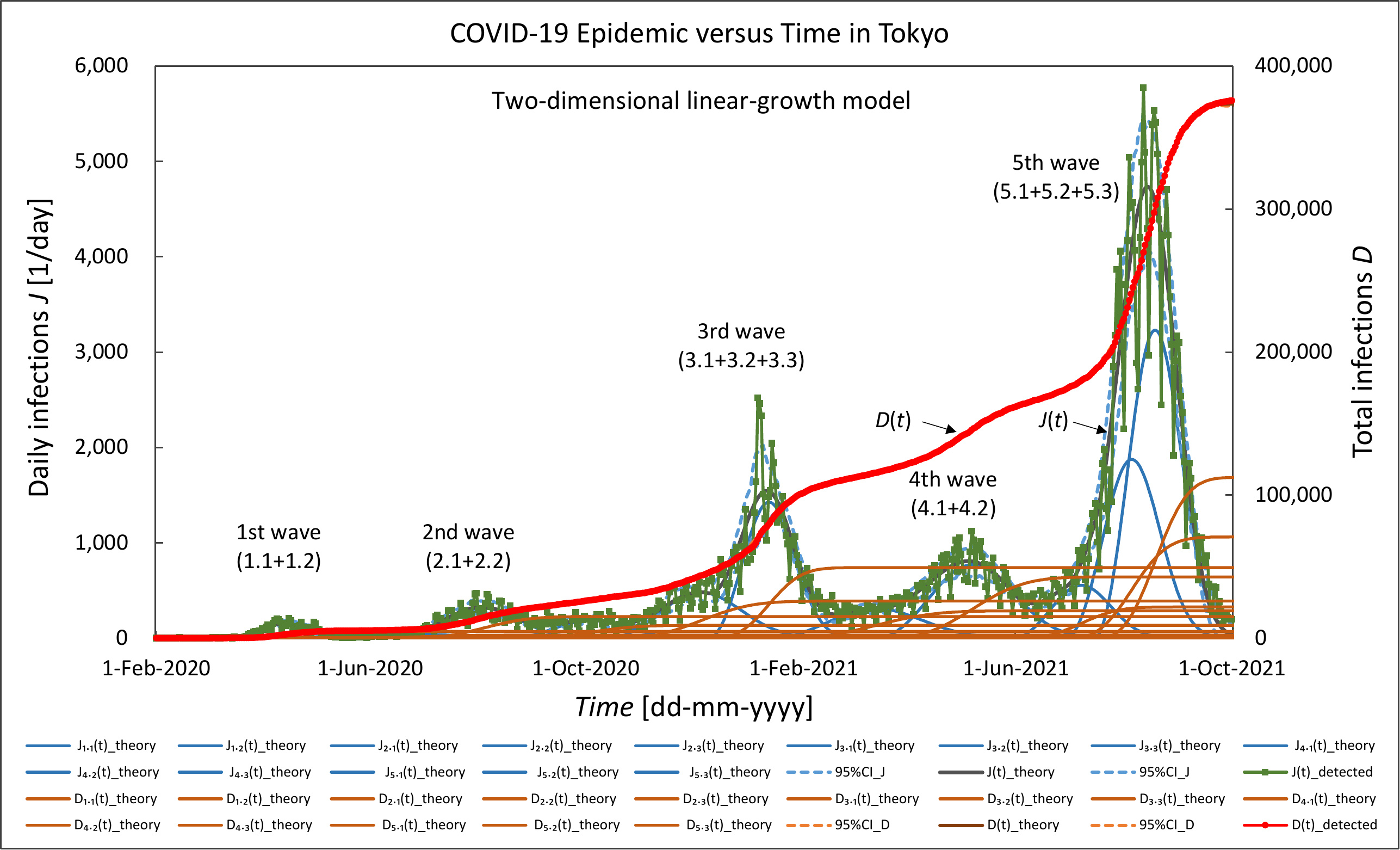} 
	\caption {Detected values and simulated results of COVID-19 infection waves over the entire duration in Tokyo. The dark green and the red markers are the detected daily and total infections, respectively. The blue curve is the theoretical value of the daily infections for each wave and the dark blue curve is the sum of them, $J(t)$. The brown curve is the theoretical value of the total infections for each wave and the dark brown curve is the sum of them, $D(t)$. The broken line is the $95\% CI$.
	} 
	\label {fig:allwaves_linear} 
\end {figure}

\begin {figure} [p] 
	\centering 
	\includegraphics[width=8.0cm]{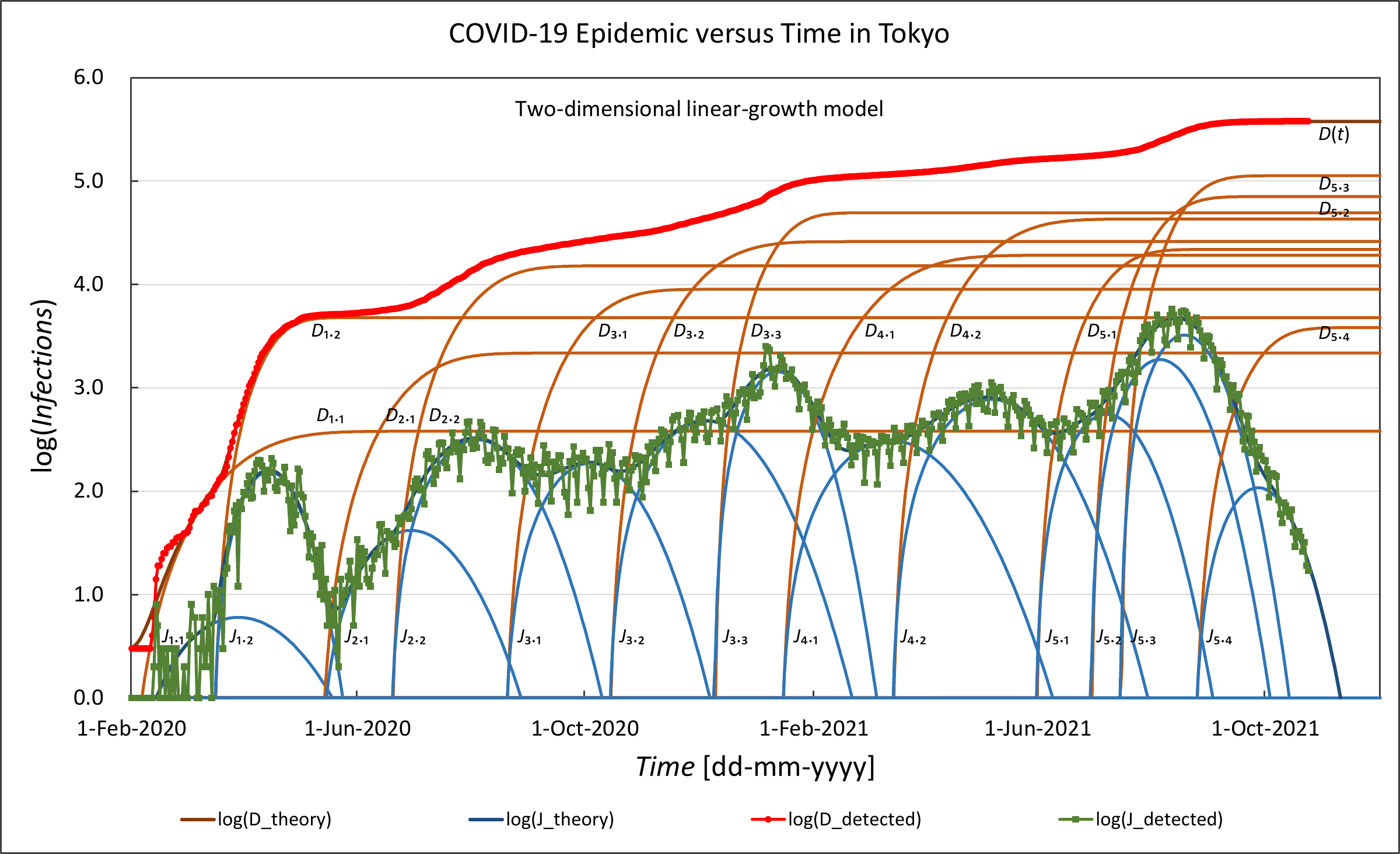} 
	\caption {Detected values and simulated results of COVID-19 infection waves over the entire duration in Tokyo. The meanings of the markers and the curves are the same as those in Fig. \ref{fig:allwaves_linear}. The entire waveform was simulated with a total of 13-wave superpositions grouped into 5 major waves.
	}
	\label {fig:allwaves_semilog} 
\end {figure}

First, the results of analysis of the entire waveform of the daily and the total infections in Tokyo are shown in Figs. \ref{fig:allwaves_linear} and \ref{fig:allwaves_semilog} as linear and semi-logarithmic plots, respectively. The dark green and the red markers are the detected daily and total infections, respectively. The blue curve is the theoretical value of the daily infections for each wave and the dark blue curve is the sum of them, $J(t)$. The brown curve is the theoretical value of the total infections for each wave and the dark brown curve is the sum of them, $D(t)$. The broken line is the $95\% CI$ calculated from the moving standard deviation with 7-days sliding window.

The entire waveform was simulated with a total of 13-wave superpositions grouped into 5 major waves. The detected value of the daily infections fluctuates significantly in a weekly cycle. This is because Sunday is a holiday in Japan and many medical institutions are also closed. On the other hand, the total infections $D(t)$ are the cumulative number of the daily infections, so the curve is smooth. The least squares method was applied to the curve of the total infections to determine the three parameters. The accuracy of the simulated total infections $D(t)$ was as good as $95\% CI / D(t) < 2\%$ after the 2nd wave when $D(t)$ exceeded 6,000.

\subsection {Superposition analysis method} 

First, the least squares method was applied to the difference between the detected and the theoretical values of the total infections $D_{1.1}(t)$ of the 1.1th wave to determine the three parameters, $D_\mathrm{s}$, $K^2$ and $\nu$. Similar to the previous time\cite{sim_covid19_1to5_japan}, the infection wave was represented as a superposition of waves, and this operation was repeated from the 1.2th wave to the 5.4th wave. Estimating the duration of each superposed wave had a significant effect on determining the values of the three parameters. Therefore, each duration was determined by repeating the following operations: Visually set the initial values, apply the least squares method to $D(t)$ to find the three parameters, draw a waveform, superpose it with the adjacent waves, and then compare it with the detected data and confirm that the error is minimized.

Table \ref{table:waves_duration} shows the duration of each wave to which the least squares method was applied.

\begin{table} 
\caption{Start day and end day of the duration of each wave to which the least squares method is applied and the duration of each major wave.} 
	\label{table:waves_duration} 
	\centering
	\begin{tabular} {cccc} 
	\hline 
	Wave & Start day & End day & Duration\\ 
	  &   &   & [day] \\ 
	\hline \hline 
	1.1th & 31-Jan-2020 & 21-Mar-2020 & \\ 
	1.2th & 15-Mar-2020 & 30-May-2020 & 121 \\ 
	\hline 
	2.1th & 10-May-2020 & 24-Jun-2020 & \\ 
	2.2th & 18-Jun-2020 & 06-Sep-2020 & 120 \\ 
	\hline 
	3.1th & 18-Aug-2020 & 26-Oct-2020 & \\ 
	3.2th & 13-Oct-2020 & 14-Dec-2020 & \\ 
	3.3th & 09-Dec-2020 & 07-Feb-2021 & 174 \\ 
	\hline 
	4.1th & 13-Jan-2021 & 21-Mar-2021 & \\ 
	4.2th & 14-Mar-2021 & 06-Jun-2021 & 145 \\ 
	\hline 
	5.1th & 30-May-2021 & 27-Jun-2021 & \\ 
	5.2th & 29-Jun-2021 & 22-Jul-2021 & \\ 
	5.3th & 15-Jul-2021 & 26-Sep-2021 & \\ 
	5.4th & 01-Oct-2021 & 24-Oct-2021 & 148 \\ 
	\hline 
	\end{tabular} 
\end{table}

\subsection {Determined parameters} 

Table \ref{table:three_parameters} shows the parameters determined for each of the 13 waves, $D_\mathrm{s}$, $K^2$, $K$, and $t_\mathrm{on}$, where $t_\mathrm{on}$ is the rise time defined as the time when the $D(t) - t$ characteristic changes from a value of 10\% to a value of 90\%, which was originally defined in electronics.
   
\begin{table} 
  \caption {Parameters determined for each of the 13 waves.} 
  \label {table:three_parameters} 
  \centering 
  \begin {tabular} {crrrr} 
	\hline 
	Wave & \multicolumn{1}{c}{$D_\mathrm{s}$} & \multicolumn{1}{c}{$K^2$} & \multicolumn{1}{c}{$K$} & \multicolumn{1}{c}{$t_\mathrm{on}$}\\ 
	& [person] & [$1/\si{day}^2$] & \multicolumn{1}{c}{[$1/\si{day}$]} & \multicolumn{1}{c}{[$\si{day}$]} \\ 
	\hline \hline 
	1.1th & 380 & 0.00212 & 0.0460 & 64 \\ 
	1.2th & 4,766 & 0.01623 & 0.1274 & 31 \\ 
	\hline 
	2.1th & 2,168 & 0.00364 & 0.0604 & 52 \\ 
	2.2th & 15,110 & 0.00453 & 0.0673 & 48 \\ 
	\hline 
	3.1th & 8,975 & 0.00455 & 0.0675 & 48 \\ 
	3.2th & 25,957 & 0.00325 & 0.0570 & 54 \\ 
	3.3th & 49,316 & 0.01154 & 0.1074 & 35 \\ 
	\hline 
	4.1th & 19,138 & 0.00225 & 0.0475 & 62 \\ 
	4.2th & 42,875 & 0.00301 & 0.0549 & 55 \\ 
	\hline
	5.1th & 21,814 & 0.00625 & 0.0894 & 40 \\ 
	5.2th & 70,764 & 0.00900 & 0.0949 & 37 \\ 
	5.3th & 112,498 & 0.01132 & 0.1064 & 35 \\
	5.4th & 3,819 & 0.01083 & 0.1041 & 36 \\
	\hline
  \end {tabular} 
\end {table}

\begin {figure} 
	\centering 
	\includegraphics[width=8.0cm] {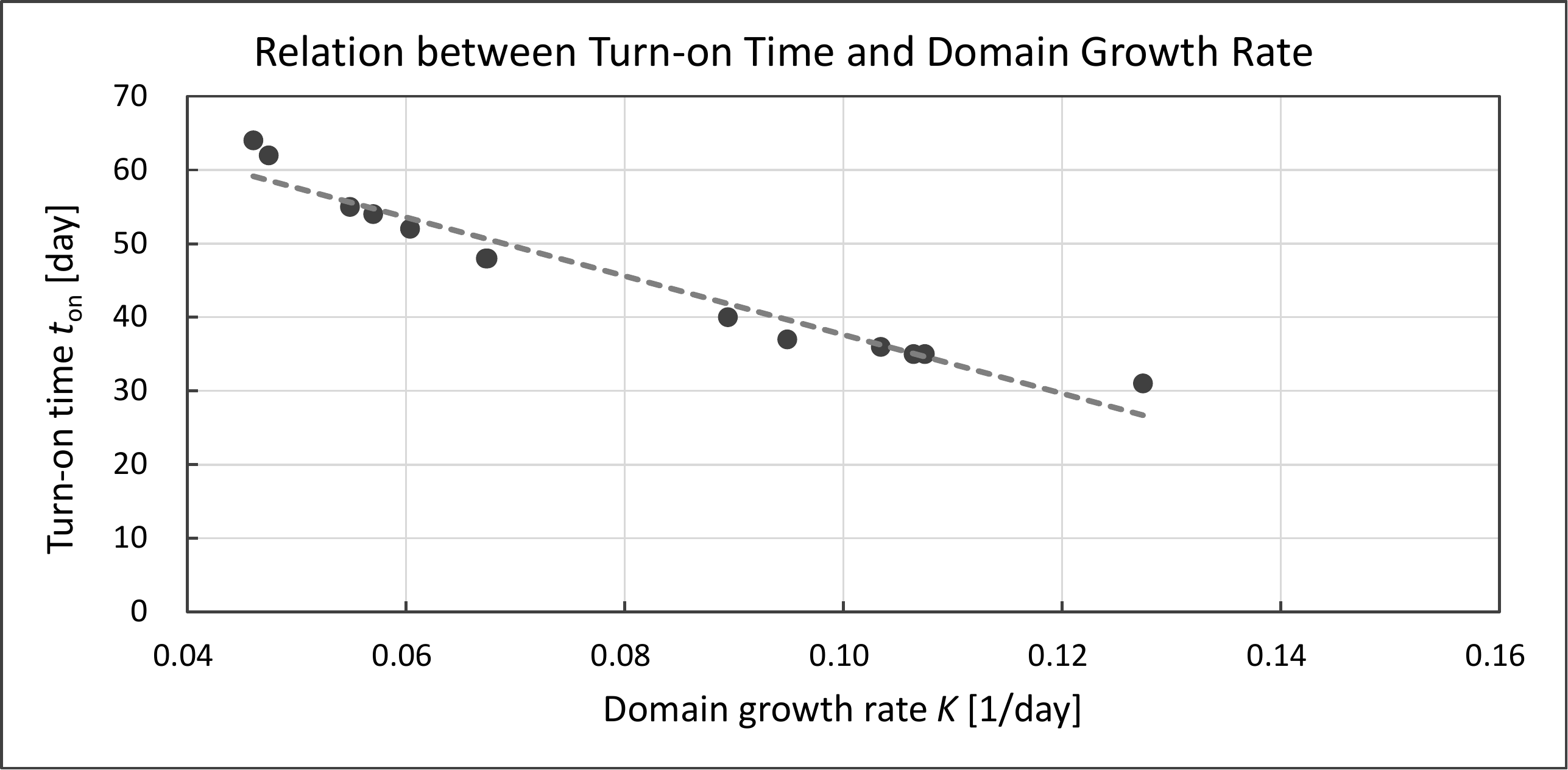} 
	\caption {Relation between the rise time $t_\mathrm{on}$ and the domain growth rate $K$, where the rise time is defined as the time taken by the $D(t) - t$ characteristic to change from the 10\% value to the 90\% value. 
	} 
	\label {fig:ton_k} 
\end {figure}

The highest correlation between each parameter was the relationship between $t_\mathrm{on}$ and $K$. The relationship is shown in Fig. \ref{fig:ton_k}. The sample correlation coefficient $\gamma$ is -0.97, and the faster the growth rate, the shorter the time it takes for the total infections to reach $D_\mathrm{s}$. This corresponds to the physical process in which the higher the domain growth rate, the faster the domains will collide with each other and reach the growth limit boundary. 

On the other hand, the value of $\gamma$ between $K$ and $D_\mathrm{s}$ was as small as 0.34. It is suggested that the domain growth rate does not significantly affect the value of $D_\mathrm{s}$. In addition, when $K$ is high and $D_\mathrm{s}$ is also large, the daily infections show a rapid increase and a rapid decrease, and the value of $t_\mathrm{on}$ is shortened.

In the previous analysis of infection waves in Japan as a whole\cite{sim_covid19_1to5_japan}, the three parameters were obtained independently. The above relationship of Tokyo is almost the same as the result of Japan.

\section{Discussions} 
\subsection{Nucleation and growth model} 

In order for one new phase to be produced in the parent phase, a new phase must be born (nucleation) and it must grow\cite{leo_mandelkern}. In the case of COVID-19, the nucleation means that new infections occur among susceptibility holders. However, unlike polarization reversal of ferroelectric materials\cite{ferro_pol_rev, nylon_pol_rev}, new infections do not occur spontaneously. As an entity model, the following physical process is assumed.

Primary infection occurs when infectious virus carriers move into a susceptibility holder population. This primary infection is considered as the nucleation. The time and location of nucleation change randomly as the virus carrier moves. If a new infection occurs at each time and place, it will spread one-dimensionally or two-dimensionally.

\subsection{The first and the second waves} 

\begin {figure} 
	\centering 
	\includegraphics [width=8.0cm] {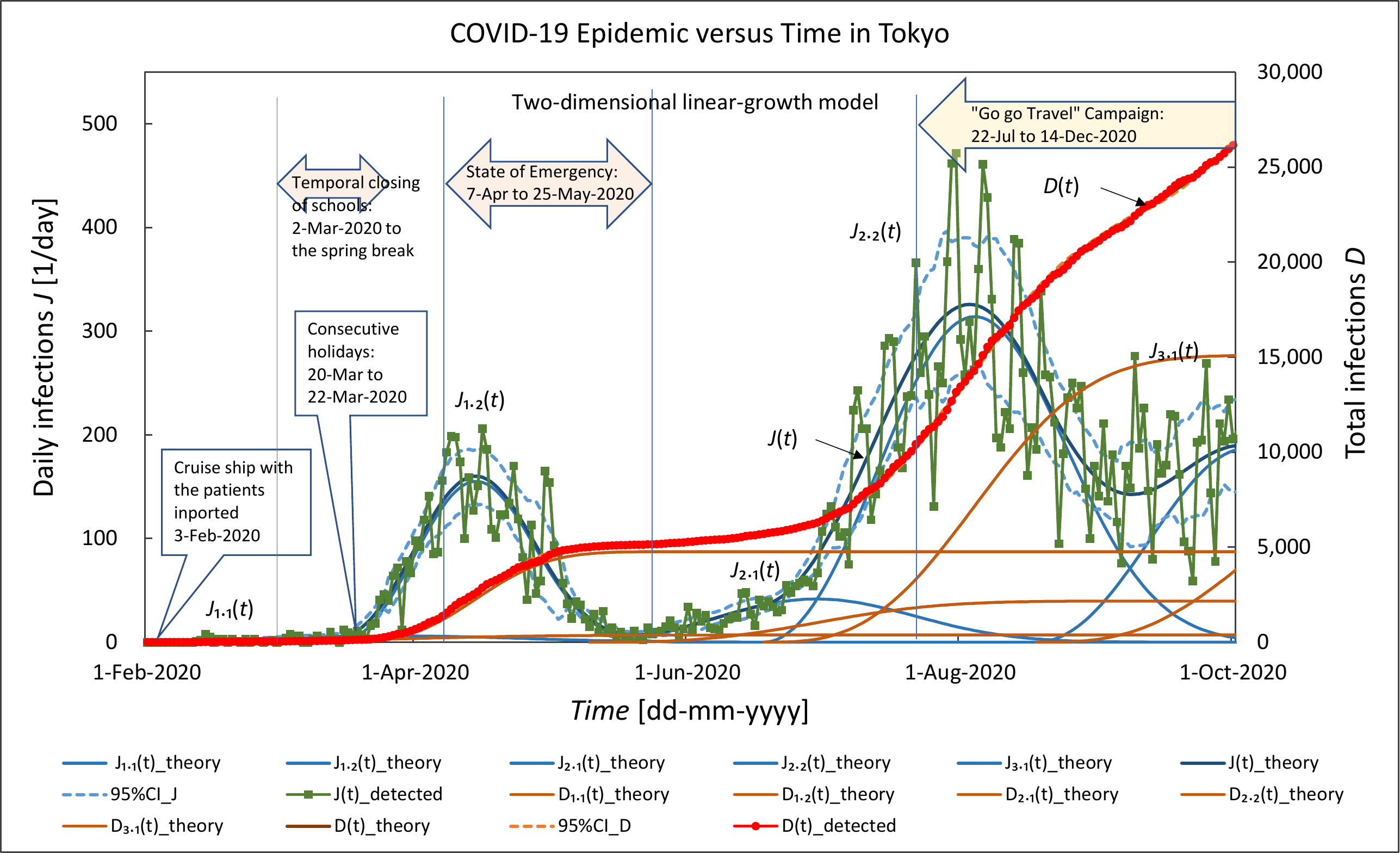} 
	\caption {Detected values and simulated results of the 1st and the 2nd COVID-19 infection waves in Tokyo. The meanings of the markers and the curves are the same as those in Fig. \ref{fig:allwaves_linear}.
	}
	\label {fig:1st_2ndwaves} 
\end {figure}

Figure \ref{fig:1st_2ndwaves} shows the characteristics of the 1st and the 2nd waves in Tokyo. The meaning of the markers and each line is the same as in Fig. \ref{fig:allwaves_linear}. The characteristics of Tokyo are similar to the characteristics of the 1st and the 2nd waves of Japan as a whole reported last time\cite{sim_covid19_1to5_japan}. The 1st wave decreased to 10 on May 25, 2020, when the state of emergency\cite{stateofemergency1} was lifted nationwide. In Japan as a whole, the number of new infections was low during June\cite{sim_covid19_1to5_japan}, but in Tokyo, as can be seen in Fig. \ref{fig:1st_2ndwaves}, it gradually increased from June and rapidly increased in early July, forming the 2nd wave. This is an example in which the spread of infection in Tokyo precedes other prefectures.

The government did not declare a state of emergency during the 2nd wave. However, the 2nd wave decreased because the inspection system was better developed than during the 1st wave and it became possible to grasp the whole picture of infected people and people refrained from acting\cite{2ndwave_kutuna}. After reaching the maximum value of the daily infections, 472 on July 31, 2020\cite{metro_tokyo}, it decreased along the theoretical curve. However, the daily infections did not decrease much, and began to increase in mid-September and continued to the 3rd wave.

\subsection{New Year holiday effect}

\begin {figure} 
	\centering 
	\includegraphics [width=8.0cm] {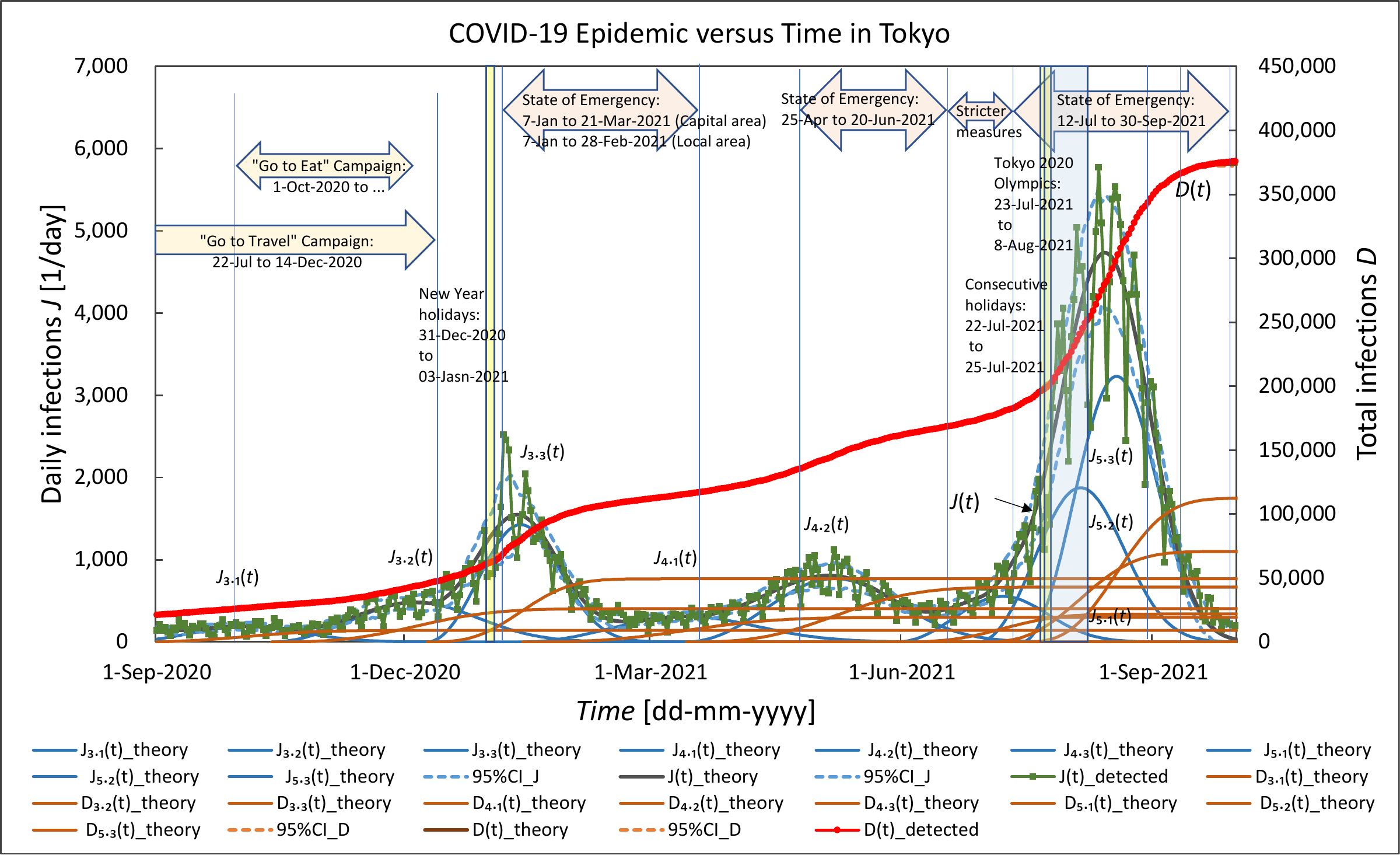} 
	\caption {Detected values and simulated results of the 3rd, the 4th, and the 5th COVID-19 infection waves in Tokyo. The meanings of the markers and the curves are the same as those in Fig. \ref{fig:allwaves_linear}.
	}
	\label {fig:3rd_5thwaves} 
\end {figure}

Figure \ref{fig:3rd_5thwaves} shows the characteristics from the 3rd to the 5th waves of COVID-19 in Tokyo. The figure shows the periods of "Go to Travel Campaign" (from 22-Jul-2020 to 14-Dec-2020) and "Go to Eat Campaign" (from 1-Oct-2020 to …) projects carried out by the government of Japan\cite{goto_campaign}, and the periods of three times of state of emergency\cite{stateofemergency2}. Of all the 5 major waves, the 3rd wave recorded the longest infection duration of 174 days (see Table \ref{table:waves_duration}).

As can be seen in Fig. \ref{fig:3rd_5thwaves}, the number of daily infections in the 3rd and 5th waves has remarkably high peaks compared with the other waves. The attribute common to these two peaks is that they appeared after the consecutive holidays. The holidays from December 31, 2020 to January 3, 2021 are New Year holidays, and as a tradition, many Japanese people move and interact nationwide.

\begin {figure} 
	\centering 
	\includegraphics [width=8.0cm] {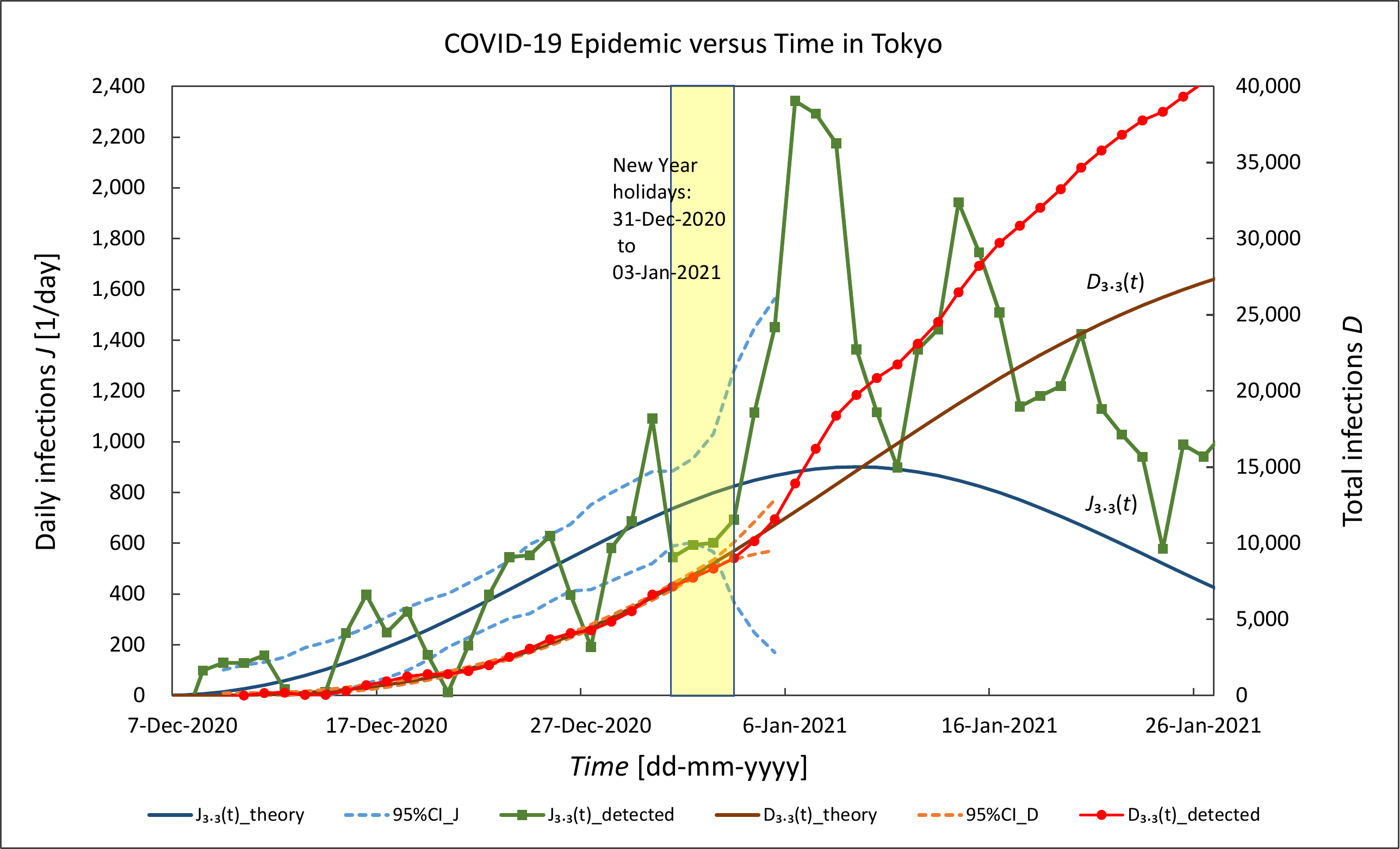} 
	\caption {Detected values and simulated results before the holidays of the 3.3th COVID-19 infection wave in Tokyo. The meanings of the markers and the curves are the same as those in Fig. \ref{fig:allwaves_linear}.
	}
	\label {fig:3rd_3wave} 
\end {figure}

Figure \ref{fig:3rd_3wave} shows the characteristics of extracted 3.3th wave from the 3rd wave shown in Fig. \ref{fig:3rd_5thwaves} by the method reported earlier\cite{sim_covid19_1to5_japan}. The thick blue and the brown curves represent the theory of the daily and the total infections, respectively, obtained from the detected infections before the holidays (until 31-Dec-2020). It is clear that the number of infections after the holidays increased significantly compared with the simulation before the holidays. 

\begin {figure} 
	\centering 
	\includegraphics [width=8.0cm] {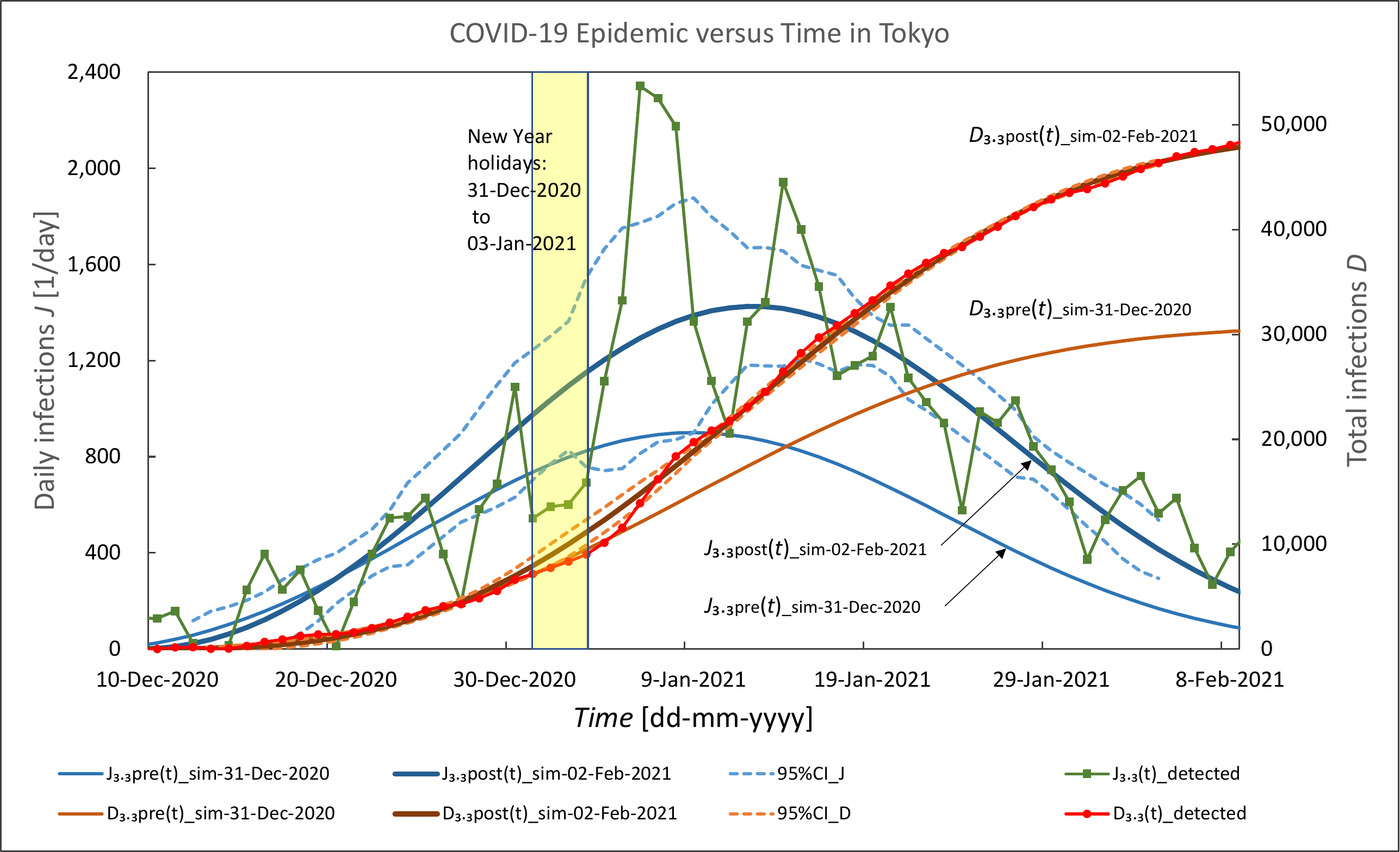} 
	\caption {Detected values and simulated results before and after the holidays of the 3.3th COVID-19 infection wave in Tokyo. The meanings of the markers and the curves are the same as those in Fig. \ref{fig:allwaves_linear}.
	}
	\label {fig:3rd_3wave_comp} 
\end {figure}

Figure \ref{fig:3rd_3wave_comp} shows the characteristics of the extracted 3.3th wave showing two simulated curves. The fluctuations in the rapid increase and decrease in the number of daily infections subsided in about 10 days, and returned to the decreasing characteristics of the one-week cycle. The figure shows the simulation characteristics performed before the consecutive holidays (until 31-Dec-2020), $J_{\mathrm{3.3pre}}(t)$ and $D_{\mathrm{3.3pre}}(t)$, and those performed after the consecutive holidays (until 02-Feb-2021), $J_{\mathrm{3.3post}}(t)$ and $D_{\mathrm{3.3post}}(t)$. The initial susceptible $D_\mathrm{s} = 49,316$ in the post-holiday simulation was 160\% of Ds = 30,808 in the pre-holidays.
This increase is thought to be due to increased interaction among people celebrating the New Year, and is therefore characterized as the "New Year holiday effect". This value is almost the same as that of Japan as a whole in 2021, 156\%\cite{sim_covid19_1to5_japan}.

\subsection{Tokyo Olympics holiday effect}

The government of Japan moved the holidays to coincide with the Tokyo 2020 Olympic Games (23-Jul-2021 to 8-Aug-2021), and made the holidays from 22th to 25th of July consecutive holidays\cite{holidaymove}.

\begin {figure} 
	\centering 
	\includegraphics [width=8.0cm] {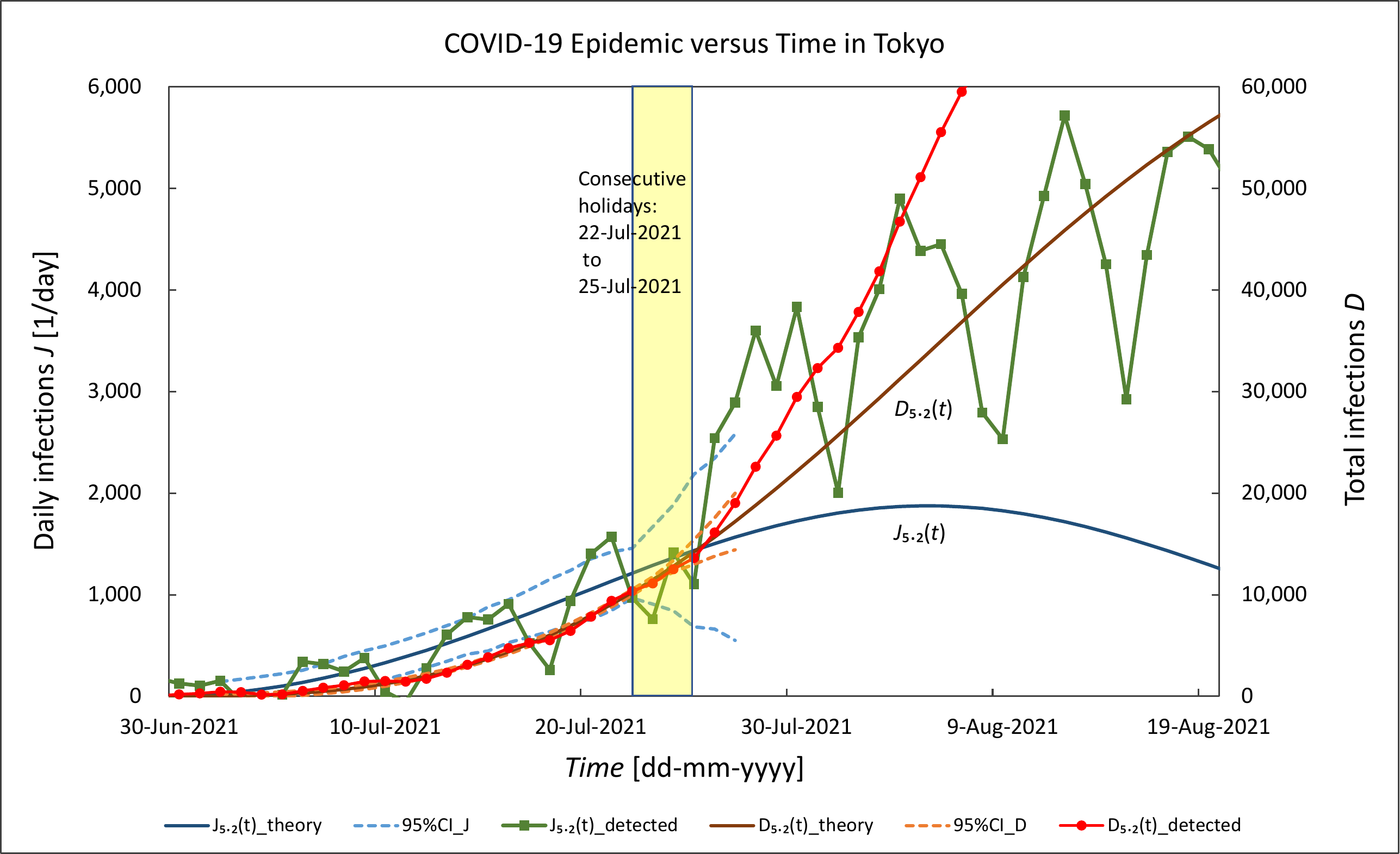} 
	\caption {Detected values and simulated results before the holidays of the 5.2th COVID-19 infection wave in Tokyo. The meanings of the markers and the curves are the same as those in Fig. \ref{fig:allwaves_linear}.
	}
	\label {fig:5th_2wave_comp} 
\end {figure}

Figure \ref {fig:5th_2wave_comp} shows the characteristics of extracted 5.2th wave from the 5th wave shown in Fig. \ref{fig:3rd_5thwaves} by the method reported earlier\cite{sim_covid19_1to5_japan}. The thick blue and the brown curves represent the theory of the daily and the total infections, respectively, obtained from the detected infections before the holidays (until 22-Jul-2021). During the consecutive holidays (22-Jul-2021 to 25-Jul-2021), the detected value is slightly smaller, but it increases remarkably immediately after the holidays.

\begin {figure} 
	\centering 
	\includegraphics [width=8.0cm] {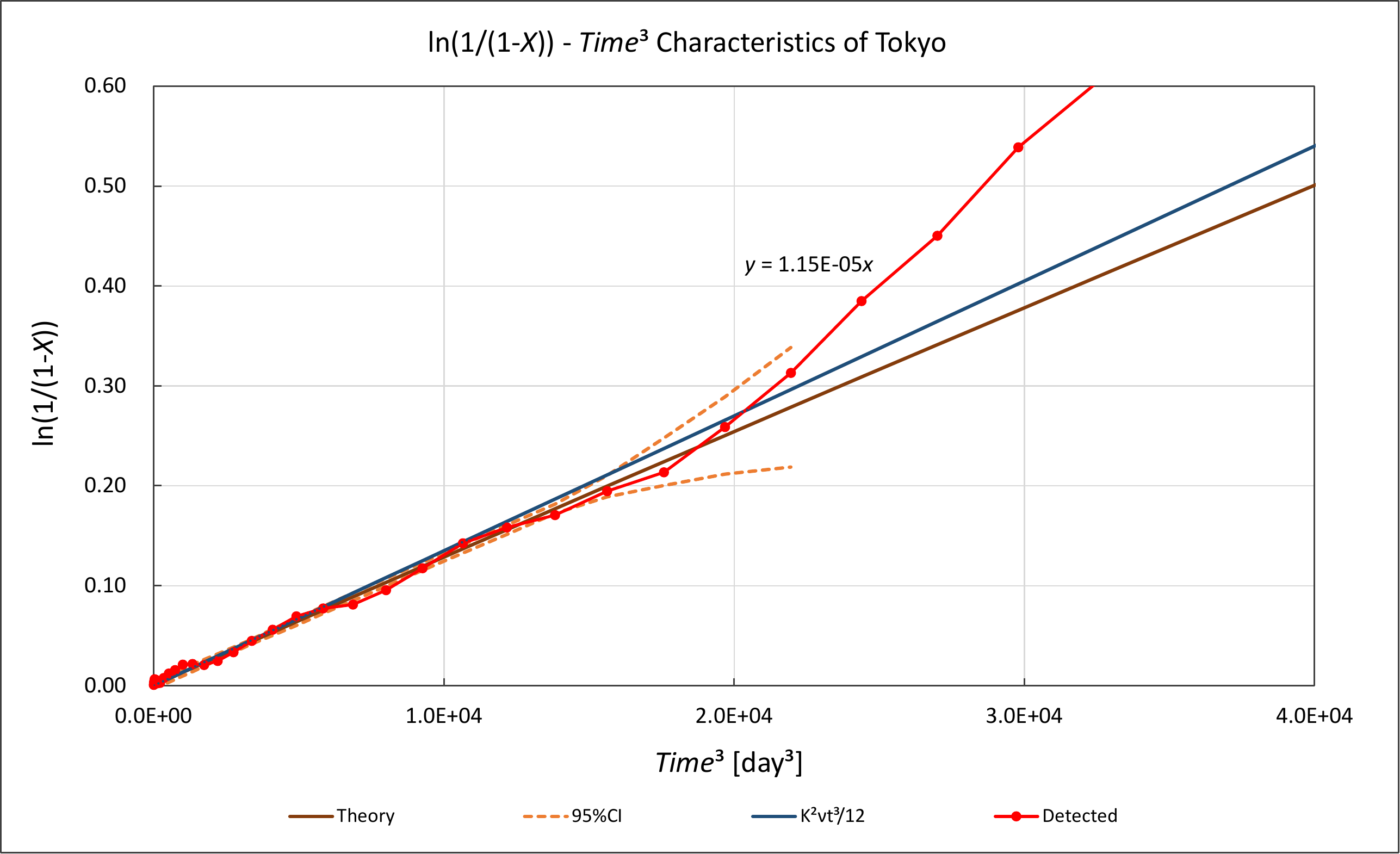} 
	\caption {The $\ln (1/(1-X)) - Time^3$ plot of the Avrami equation for the 5.2th wave. $X$ is the fraction of domain that has been transformed by formation and growth of nuclei. The red marker is the detected value, the dark brown line represents the theory, the dark blue straight line is the approximate straight line when $\nu t$ is small enough, and the broken line represents the $95\% CI$.
	} 
	\label {fig:y_t3_5th_2wave} 
\end {figure}

This feature is more clearly shown in Fig. \ref {fig:y_t3_5th_2wave} as a deviation from the straight line of the $\ln (1 / (1-X)) – Time ^ {3}$ plot of the Avrami equation\cite{leo_mandelkern}, where $X$ is the fraction of domain that has been transformed by formation and growth of nuclei.

\begin {figure} 
	\centering 
	\includegraphics [width=8.0cm] {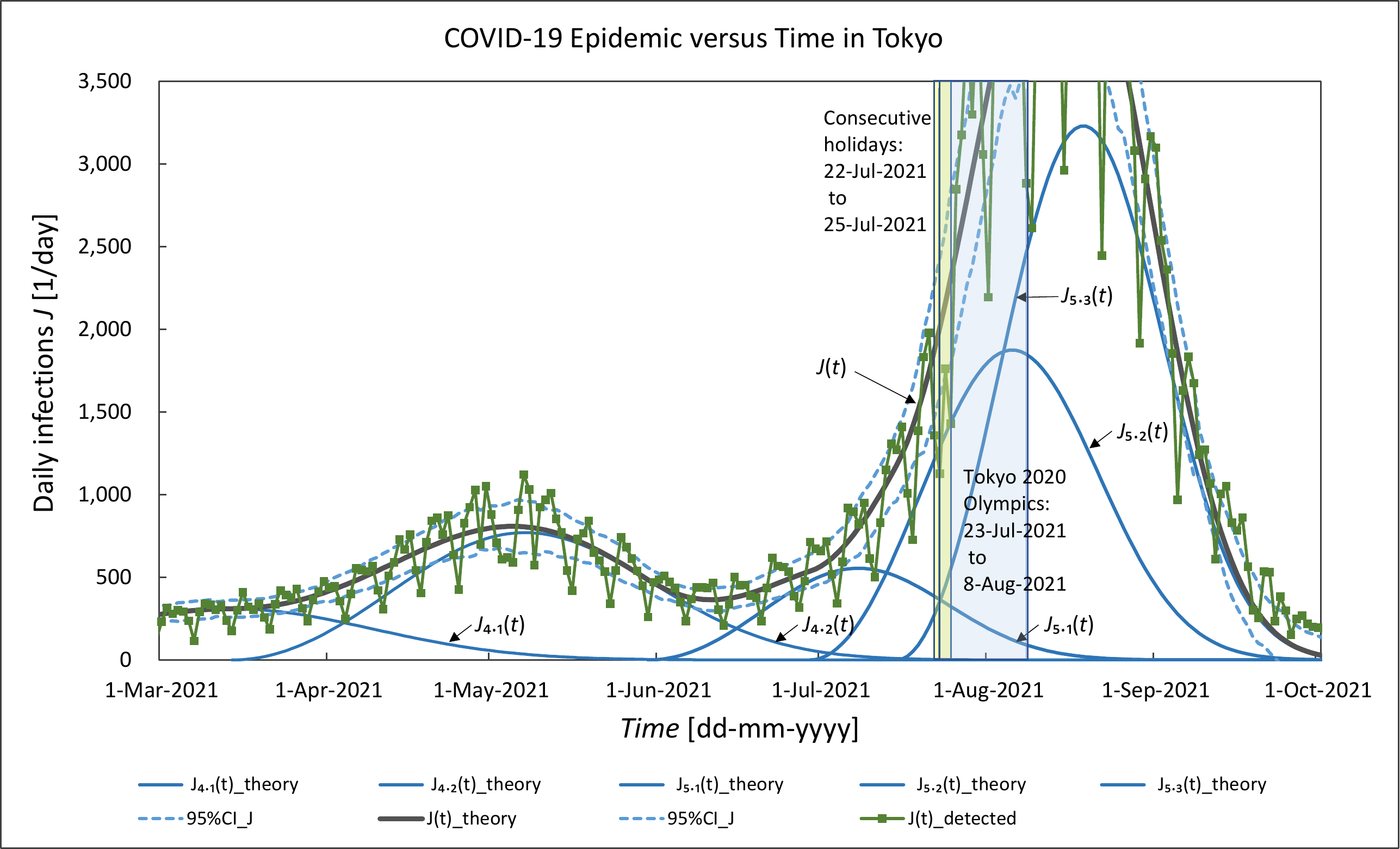} 
	\caption {Detected values and simulated results of the 4th and the 5th COVID-19 infection waves in Tokyo. The meanings of the markers and the curves are the same as those in Fig. \ref{fig:allwaves_linear}.
	}
	\label {fig:4and5wave} 
\end {figure}

The characteristics in Fig. \ref {fig:5th_2wave_comp} are similar to those for the New Year holidays in Fig. \ref{fig:3rd_3wave}, but the characteristics after the holidays are different. The rapid fluctuation in the number of daily infections after the New Year holidays subsided in about 10 days, and returned to the decreasing characteristic of the one-week cycle. However, as can be seen in Fig. \ref {fig:5th_2wave_comp}, the number of infections after the consecutive holidays of the Tokyo Olympics continued to increase even after 10 days, so it was regarded as the occurrence of the new 5.3th wave, as shown in detail in Fig. \ref {fig:4and5wave}.

\begin {figure} 
	\centering 
	\includegraphics [width=8.0cm] {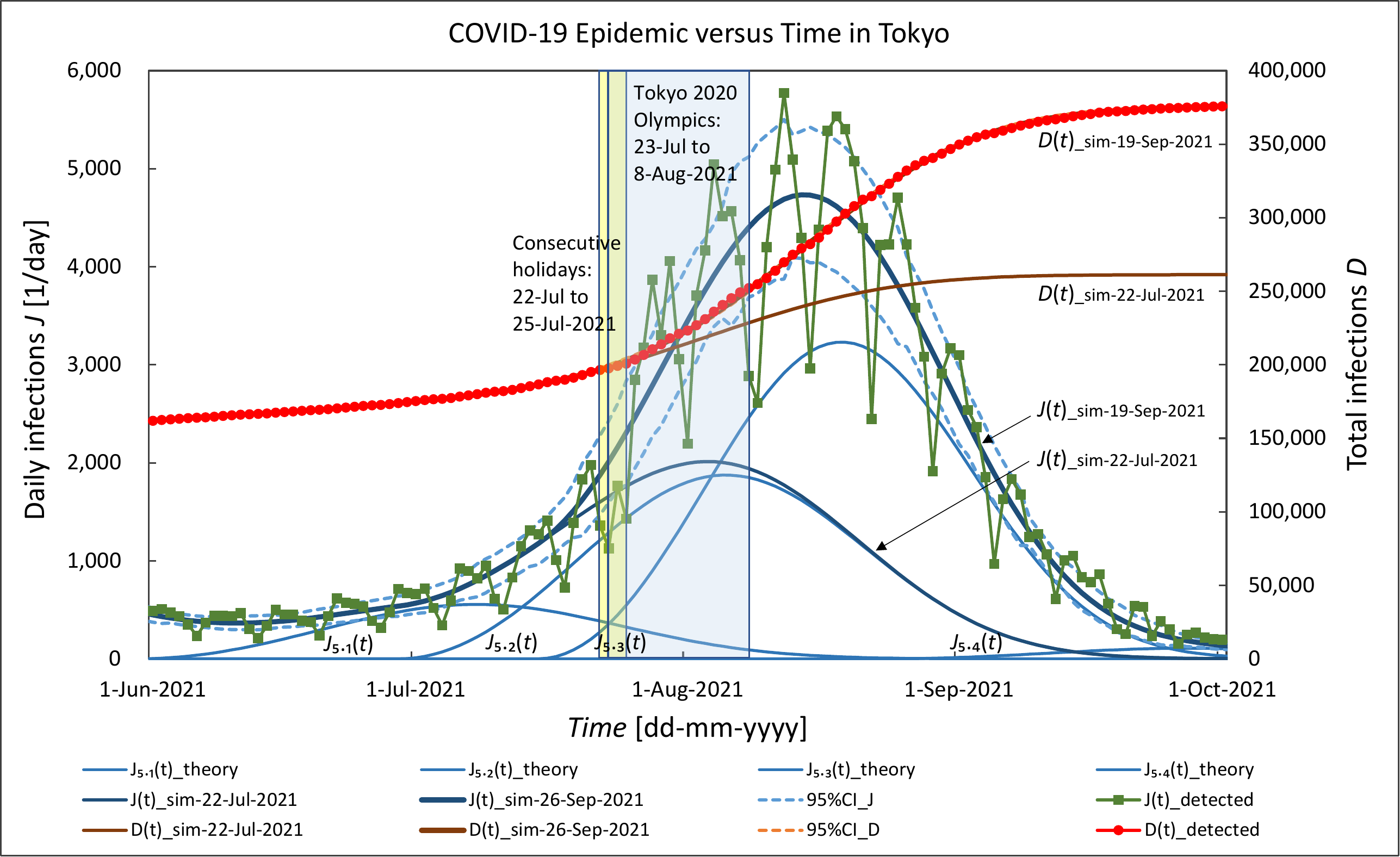} 
	\caption {Detected values and simulated results before and after the holidays of the 5th COVID-19 infection wave in Tokyo. The meanings of the markers and the curves are the same as those in Fig. \ref{fig:allwaves_linear}.
	}
	\label {fig:inf_t_5wave} 
\end {figure}

Figure \ref {fig:inf_t_5wave}  shows all the characteristics of the 5th wave. The figure shows the simulation characteristics performed before the consecutive holidays (until 22-Jul-2021), $J_{\mathrm{pre}}(t)$ and $D_{\mathrm{pre}}(t)$, and those performed after the consecutive holidays (until 26-Sep-2021), $J_{\mathrm{post}}(t)$ and $D_{\mathrm{post}}(t)$.

In the pre-holiday simulation, the initial susceptible was $D_\mathrm{s} = D_{s5.1} + D_{s5.2} = 92,578$, but in the post-holiday simulation, it was $D_\mathrm{s} = D_{s5.1} + D_{s5.2} + D_{s5.3} + D_{s5.4} = 208,895$. The value for the post-holiday was about 226\% of that for the pre-holiday. This increase is due to the 5.3th wave. It occurred during the consecutive holidays and increased remarkably during the Tokyo 2020 Olympic Games. Within two weeks thereafter, the 5th wave recorded the highest number of daily infections ever (5,773 on 12-Aug-2021 and 5,534 on 18-Aug-2021\cite{metro_tokyo}). By the time of the 5.3th wave, the replacement of the virus with the more infectious Delta variant was almost completed, as described in the next subsection. 

It is considered to be the essential condition for the 5.3th wave that the Olympics were held (without spectators, though), and the necessary conditions that the consecutive holidays were set to celebrate the Olympics, the diverse exchanges of people increased in celebration of the Olympics, and the Delta variant caused more infectious nucleation in various places than ever before (the growth rate $K$ of the 5.3th wave did not change much from the 5.2th wave, see Table \ref{table:three_parameters}). The increase in the number of infections caused by the 5.3th wave is characterized as the "Tokyo Olympics holiday effect" separately from the "Delta variant effect" described in the next subsection.

\subsection{Delta variant effect}

\begin {figure} 
	\centering 
	\includegraphics [width=8.0cm] {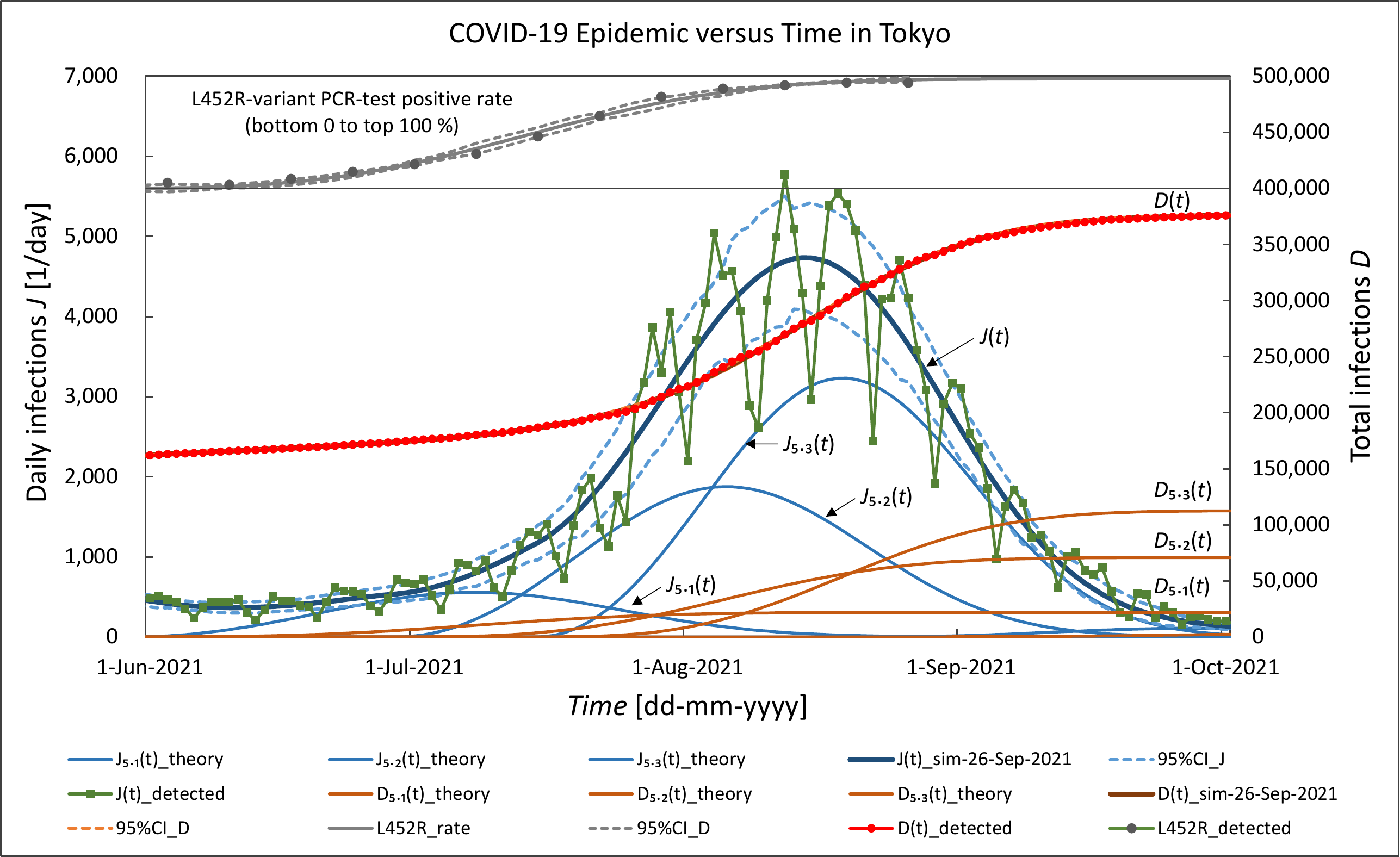} 
	\caption {Detected values and simulated results of the 5th COVID-19 infection wave and the L452R-variant PCR-test result in Tokyo. The meanings of the markers and the curves are the same as those in Fig. \ref{fig:allwaves_linear}.
	}
	\label {fig:inf_l452r_t_5wave} 
\end {figure}

In Japan, during the 5th wave, the major virus of Alpha variant type was replaced by the more infectious Delta variant. Since the test result L452R-variant PCR-test positive rate conducted in Tokyo has been published\cite{l452r_screening}, the analysis result of the 5th wave in Tokyo with that data inserted at the top of the graph is shown in Fig. \ref {fig:inf_l452r_t_5wave}.

\begin {figure} 
	\centering 
	\includegraphics [width=8.0cm] {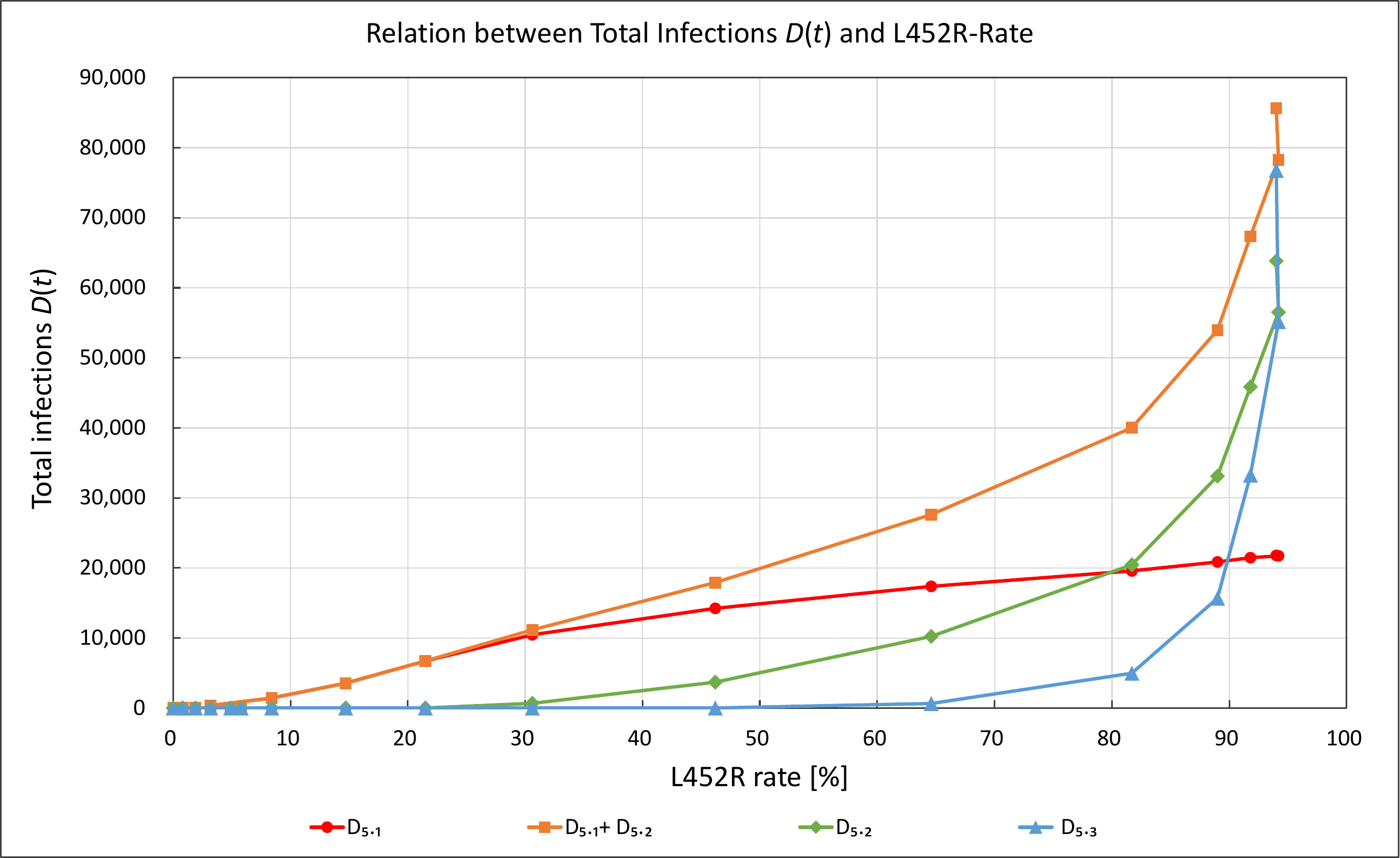} 
	\caption {Detected total infections, $D_{5.1}(t)$, $D_{5.1}(t) + D_{5.2}(t)$, $D_{5.2}(t)$, and $D_{5.3}(t)$ as a function of the L452R rate in Tokyo.
	}
	\label {fig:d_l452r_5wave} 
\end {figure}

As can be seen in Fig. \ref {fig:inf_l452r_t_5wave}, the curves of the L452R rate and the curves of $D_{5.1}(t)$, $D_{5.2}(t)$, and $D_{5.3}(t)$ are similar. Then, $D_{5.1}(t)$, $D_{5.1}(t) + D_{5.2}(t)$, $D_{5.2}(t)$, and $D_{5.3}(t)$ from April 29, 2021 to August 26, 2021 are plotted as a function of the L452R rate, and the characteristics are shown in Fig. \ref {fig:d_l452r_5wave}. 

\begin {figure} 
	\centering 
	\includegraphics [width=8.0cm] {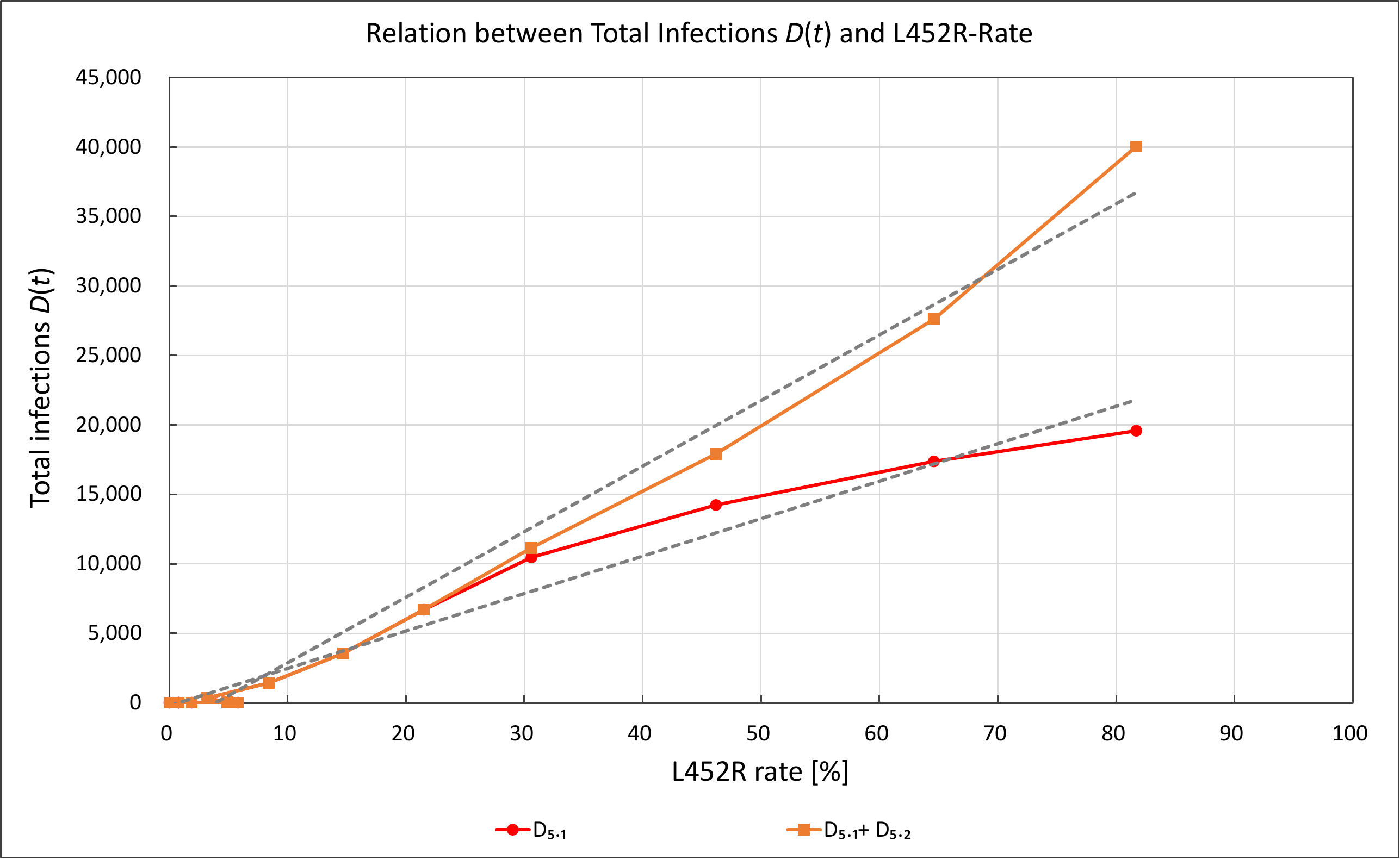} 
	\caption {Detected total infections, $D_{5.1}(t)$ and $D_{5.1}(t) + D_{5.2}(t)$ as a function of the L452R rate in Tokyo.
	}
	\label {fig:d1_l452r_5wave} 
\end {figure}

Since $D_{5.1}(t)$ is close to a straight line for the whole period and $D_{5.1}(t) + D_{5.2}(t)$ is close to a straight line for the period until July 29, 2021, their characteristics are shown in Fig. \ref {fig:d1_l452r_5wave}. Both show almost linear relationships, and the sample correlation coefficients are 0.985 and 0.992, respectively. Since this quantitative evaluation targets only the virus, the high correlation is considered to represent causality rather than mere correlation. In the period from April 29, 2021 to July 29, 2021, the L452R rate reached about 82\%. After August, the L452R rate increased by about 10\%, but the total infections increased more rapidly than that, as seen in the characteristics of $D_{5.2} (t)$ and $D_{5.3} (t)$ in Fig. \ref {fig:d_l452r_5wave}. 

In addition, the domain growth rates $K$ of the 5.1th wave, the 5.2th wave, and the 5.3th wave are about 163\%, 173\%, and 194\%, respectively, compared with the value of the 4.2th wave which was caused by the Alpha variant (see Table \ref{table:three_parameters}).

From the above analysis, it was concluded that the cause of the 5.1th and the 5.2th waves was the replacement of the Alpha variant with the Delta variant.
The $D_\mathrm{s}$ = 92,578 of the 5.1th wave + 5.2th wave is about 150\% of the $D_\mathrm{s}$ = 62,014 of the 4th wave which was caused by the Alpha variant.
In addition, the $D_\mathrm{s}$ = 208,895 of the 5th wave by almost the Delta variant is about 248\% of the $D_\mathrm{s}$ = 84,249 of the 3rd wave which is the largest ever by the Alpha variant. The doubling of these infections is characterized as the "Delta variant effect".

\subsection{Predictability of infections}

The 5.3th wave was analyzed to see the predictability of infections in the near future and the vaccination effects. Table \ref {table:predictability} shows the results when the end day of the 5.3th wave in Table \ref{table:waves_duration} is changed.

The value of $D_\mathrm{s}$ obtained in the simulation was 61,958 when the duration was until August 10, 2021 two days after the Olympics final day, but it increased significantly when the duration was until August 15, 2021. After that, it became an almost constant value (about 110,000). The total infections in the near future could be predicted within ± 2\% from the detected values after August 29, 2021 when the number of daily infections peaked. 

\subsection{Vaccination effect}

The number of daily infections in the 5th wave decreased to a small value (17 on 24-Oct-2021\cite{metro_tokyo}) next to the 1st wave (10 on 25-May-2020). 
The period between each component of the 5 major waves (e.g. $t(J_{2.1max}) – t(J_{1.1max})$, etc.) averaged about 120 days. Predicting from this cycle, it is natural that the initial wave of the 6th wave will occur in late October ($J_{6.1max}$ could occur on 2-Nov-2021). From the end of September, the detected values tended to increase slightly from the theoretical values. This component was small and did not tend to increase even in October, so it was regarded as the 5.4th wave (see Fig. \ref {fig:allwaves_semilog}). The $D_\mathrm{s}$ of the 5.4th wave is 43\%, 20\%, and 18\% of the 3.1th wave, the 4.1th wave, and the 5.1th wave, respectively (see Table \ref {table:three_parameters}). The fact that the 5.4th wave that could be the initial wave of the 6th wave is suppressed to about 20\% to 40\% of the initial waves of the past major waves is considered to be the vaccination effect.

\begin{table} [h] 
  \caption {The predictability simulation.} 
  \label {table:predictability} 
  \centering 
  \begin {tabular} {crrr} 
	\hline 
	End day & \multicolumn{1}{c}{$D_\mathrm{s}$} & \multicolumn{1}{c}{$K^2$} & \multicolumn{1}{c}{$D_\mathrm{s}$ ratio}\\ 
	of simulation & [person] & [$1/\si{day}^2$] & \multicolumn{1}{c}{[$\si{\%}$]} \\ 
	\hline \hline 
	10-Aug-2021 &   61,958 &  0.02380 &  55.1 \\ 
	15-Aug-2021 & 107,330 &  0.01216 &  95.4 \\  
	22-Aug-2021 & 106,486 &  0.01229 &  94.7 \\ 
	29-Aug-2021 & 111,395 &  0.01158 &  99.0 \\ 
	05-Sep-2021 & 111,276 &  0.01159 &  98.9 \\ 
	12-Sep-2021 & 111,014 &  0.01164 &  98.7 \\ 
	19-Sep-2021 & 111,730 &  0.01149 &  99.3 \\ 
	26-Sep-2021 & 112,498 &  0.01132 & 100.0 \\ 
	03-Oct-2021 & 113,187 &  0.01117 & 100.6 \\ 
	17-Oct-2021 & 114,194 &  0.01094 & 101.5 \\
	\hline
  \end {tabular} 
\end {table}

\section {Conclusion}

The purpose of this study was to simulate all COVID-19 infection waves in Tokyo, the capital of Japan, by the phase transformation dynamics theory, and to quantitatively analyze the detailed structure of the waveform for estimating the cause. The whole infection wave in Tokyo was basically expressed by the superposition of the 5 major waves as in Japan as a whole. Among these waves, the detailed structure was seen in the 3rd and the 5th waves, where the number of infections increased remarkably due to the holidays. Letting the simulated initial susceptible be $D_\mathrm{s}$, $D_\mathrm{s}$(post New Year holidays)$ / D_\mathrm{s}$(pre New Year holidays)$ > 1$ (e.g. 1.6), $D_\mathrm{s}$(post Olympics holidays)$ / D_\mathrm{s}$(pre Olympics holidays)$ > 1$ (e.g. 2.3), and $D_\mathrm{s}$(Delta variant)$ / D_\mathrm{s}$(Alpha variant)$ > 1$ (e.g. 1.5, 2.5) were characterized as the “New Year holiday effect”, the “Tokyo Olympics holiday effect”, and the “Delta variant effect”, respectively. Since this method had high simulation accuracy for the cumulative number of infections, it was effective in estimating the cause, the number of infections in the near future and the vaccination effect by quantitative analysis of the detailed structure of the waveform. Japan consists of long archipelago from north to south. Analyzing the infection waves of multiple prefectures far from the central part and comparing them with the results of Tokyo is a topic for future study.

\end{document}